\documentclass{aastex62}
\usepackage{amsmath}
\graphicspath{{./}{figures/}}

\received{March 21, 2019}
\accepted{July 1, 2019}
\submitjournal{ApJ}

\begin{document}

\title{Reduced Order Estimation of the Speckle Electric Field History for
Space-Based Coronagraphs}

\correspondingauthor{Leonid Pogorelyuk}
\email{leonidp@princeton.edu}

\author{Leonid Pogorelyuk}
\affiliation{Department of Mechanical and Aerospace Engineering \\
Princeton University \\
41 Olden St, Princeton, NJ 08544, USA}

\author{N. Jeremy Kasdin}
\affiliation{Department of Mechanical and Aerospace Engineering \\
Princeton University \\
41 Olden St, Princeton, NJ 08544, USA}

\author{Clarence W. Rowley}
\affiliation{Department of Mechanical and Aerospace Engineering \\
Princeton University \\
41 Olden St, Princeton, NJ 08544, USA}



\begin{abstract}

In high-contrast space-based coronagraphs, one of the main limiting
factors for imaging the dimmest exoplanets is the time varying nature
of the residual starlight (speckles). Modern methods try to differentiate between
the intensities of starlight and other sources, but none incorporate models of space-based systems which can take into
account actuations of the deformable mirrors. Instead, we propose
formulating the estimation problem in terms of the electric field
while allowing for dithering of the deformable mirrors. Our reduced-order
approach is similar to intensity-based PCA (e.g. KLIP) although, under
certain assumptions, it requires a considerably lower number of modes
of the electric field. We illustrate this by a FALCO simulation of the
WFIRST hybrid Lyot coronagraph.\footnote{The data and the comparison code are available at: \url{https://github.com/leonidprinceton/EFOR}}

\end{abstract}



\section{Introduction} 

Direct imaging of exoplanets and circumstellar disks requires an observatory
that creates a high contrast between the astrophysical source and
the residual starlight halo (speckles). The contrast needed to detect and
characterize Earth-like planets is on the order of $10^{10}$ and
can be achieved only by going to space to avoid the fundamental limitations
imposed by atmospheric turbulence (\cite{guyon2005limits,cavarroc2006fundamental}).
The Coronagraph Instrument
(CGI) on NASA's upcoming Wide-Field Infra-Red Survey Telescope (WFIRST) is designed to demonstrate the
viability of such missions (\cite{douglas2018wfirst}). It will employ
passive and active starlight suppression systems (coronagraphs, deformable
mirrors, etc.) to block most of the starlight and control the diffraction
pattern in a region of the image where the planet is located (\cite{demers2015requirements,demers2018review}).
Yet, the speckle background is expected to remain brighter than the
planet and to vary over the lengthy observation time (\cite{shaklan2011stability,ygouf2016data}). It is therefore necessary to post-process the images to
distinguish between the speckles and the signals incoherent with the
star (which include the planet). We will refer to the ratio between the intensity of the speckle floor and the dimmest detectable planet as the post-processing factor.

Post-processing methods for ground based telescopes exploit various
sources of diversity between images to extract the faint signal from
the varying background (for a review, see \cite{pueyo2018direct,jovanovic2018review}).
Some utilize physical properties such as the rotation of the telescope
(\cite{marois2006angular}), phase apodization (\cite{codona2006high,kenworthy2007first}), the spectrum of the planet and the star (\cite{marois2000efficient,sparks2002imaging,martinache2010kernel}),
and their polarization (\cite{baba2003method}). Others rely on mathematical
manipulations of multiple images, either only from the target star (\cite{lafreniere2007new}),
or from additional observations of reference stars (\cite{mawet2012review}),
and even artificial speckles (\cite{jovanovic2015artificial}). These
images are often used to construct a low dimensional representation
of the speckle intensity by means of Principal Component Analysis
(PCA) (\cite{soummer2012detection,amara2012pynpoint,fergus2014s4}),
Robust PCA (\cite{gonzalez2016low}), or Non-negative Matrix Factorization (\cite{ren2018non}).

However, in order to fully capture the contribution of the speckles in each image, one has to take into account the evolution of the speckles and how they are affected by the deformable mirrors (DMs). For ground based telescopes this can be done by monitoring the speckles in the coronagraph plane after being partially suppressed by the adaptive optics (Phase Sorting Interferometry, \cite{codona2013focal}). However, space based telescopes lack an adaptive optics system and operate with much lower photon counts ($\sim10^3$ photons per exposure compared to $\sim10^5$ in \cite{codona2013focal}). The phase diversity therefore has to come from dithering the DMs beyond what is necessary to suppress the speckles (as was showns by the authors in \cite{pogorelyuk2019dark}).

Although DM dithering may lead to a higher post-processing
factor, it requires accounting for the DM
control history which can be achieved by formulating the estimation
problem in terms of the electric field of the speckles rather than their intensity.
In this paper, we extend our previous work by introducing a reduced-order
representation of the speckle field, similar to the the low-rank
treatment of speckle intensities by PCA-based methods. 

We also propose that in the context of space-based coronography, formulating
the estimation problem in terms of the electric field is more natural
than in terms of the intensity. Although our method \emph{requires}
introducing phase diversity via DM dithering, it is  able
to distinguish between the incoherent signal and the residual starlight
in the absence of reference libraries, telescope rotations, spectrographs,
or other means to achieve diversity that require multiple images.

In section~\ref{sec:EFOR}, we describe an electric field order reduction
approach to the estimation of speckle history and incoherent intensity
and propose an explanation for its better performance (compared to
intensity-based estimation). The suggested algorithm incorporates
the DM commands which may come from those used for high-contrast maintenance during the observation
phase (i.e., the closed-loop approach, see \cite{miller2017spatial,pogorelyuk2019dark}) or from dithering the DMs to increase the phase diversity necessary for estimating the electric fields (an open-loop approach) or both together. In section~\ref{sec:simulation}, we use a simulated
WFIRST observation scenario to illustrate the new algorithm's performance and compare
it to an intensity-PCA-based estimation\footnote{The data and the comparison code are available at: \url{https://github.com/leonidprinceton/EFOR}}.

\section{\label{sec:EFOR}Electric Field Order Reduction}

In this section, we formulate an algorithm which computes an estimate
of the incoherent intensity (of light sources other than the star)
given measurements of photon counts across multiple pixels and time
frames and the history of DM controls. It is similar to PCA-based
methods such as Karhunen-Lo\`eve Image Projection (KLIP) (\cite{soummer2012detection})
and PynPoint (\cite{amara2012pynpoint}) since it considers a low-dimensional
basis for the speckles, but is also similar to the \emph{a posteriori} estimation method suggested in \cite{pogorelyuk2019dark} since it
formulates an optimization problem in terms of the electric field.

PCA based methods assume that the speckles lie in an empirically determined
subspace(s) of intensities.
Besides ignoring the history of control inputs and allowing for negative
intensities, these methods might also increase the dimension of the subspace
beyond what is necessary for accurately approximating the speckles in a
space-based telescope. Indeed, let us assume that the speckles drift due to slowly varying wavefront errors (WFEs) which can be well
approximated by a small number ($\sim20$) of 2D polynomials (e.g.,
Zernikes). Then, due to the linear nature of Fourier optics, the variations
in the electric field of the speckles are also well approximated by the
same small number of electric field modes. However, as we show next,
the number of intensity modes necessary to capture the same variations
is significantly larger ($\sim200$). Under these assumptions, a low
dimensional representation of the electric field reduces the number
of model parameters by an order of magnitude (while also allowing the incorporation of
the history of DM dither and controls).

Throughout the paper we consider a set of $N$ low-bandwidth photon detectors located in a region of high contrast --- the dark hole. Broadband imaging is therefore modeled by multiple detectors of discrete wavelengths and their exact number and location does not play a role in the analysis below (alternatively, one can think of taking multiple images with low-bandwith filters of varying wavelength, see for example \cite{give2007broadband}). We define the electric field vector over the real numbers (to make the discussion of the numerical method and the comparison to intensity PCA easier later in the text), as
\begin{equation}
\mathbf{E}(t)=\begin{bmatrix}\mathrm{Re}E_{1}(t)\\
\mathrm{Im}E_{1}(t)\\
\mathrm{Re}E_{2}(t)\\
\mathrm{Im}E_{2}(t)\\
\vdots
\end{bmatrix}\in\mathbb{R}^{2N},
\end{equation}
where $E_{i}(t)$ is the electric field of the residual start light
(speckles) at pixel $i$ and time frame $t$. Here $t=0$ correspond
to the first exposure after the dark hole has been created (for example,
via Electric Field Conjugation - EFC  (\cite{give2007electric}). We take the
nominal DM setting used to create the dark hole to be, without loss of
generality, $\mathbf{u}(0)=\mathbf{0}$. By exploiting the approximately
linear effects of DM actuations, we further split the electric field
into
\begin{equation}
\mathbf{E}(t)\approx\mathbf{E}^{OL}(t)+G^{U}\mathbf{u}(t),\label{eq:EOL_def}
\end{equation}
where $\mathbf{E}^{OL}(t)$ is the hypothetical speckle field in the absence
of DM actuations (open-loop), $\mathbf{u}(t)$ is the deviation of
the DM command from its nominal value (due to both dither and possible closed-loop control),
and $G^{U}$ is the control influence matrix (the Jacobian).

The intensity of the speckles is given by
\begin{equation}
\mathbf{I}^{S}(t)=B\cdot\left(\mathbf{E}(t)\circ\mathbf{E}(t)\right)\in\mathbb{R}^{N},\label{eq:speckles_intensity}
\end{equation}
where $\circ$ stands for the Hadamard (element-wise) product and
\begin{equation}
B=\begin{bmatrix}1 & 1\\
 &  & 1 & 1\\
 &  &  &  & \ddots & \ddots
\end{bmatrix}=I_{N\times N}\otimes\begin{bmatrix}1 & 1\end{bmatrix}.
\end{equation}
We define the total intensity in the image (from all sources) as
\begin{equation}
\mathbf{I}(t)=\mathbf{I}^{S}(t)+\mathbf{I}^{P}+\mathbf{I}^{Z}+\mathbf{I}^{D}\label{eq:I_def}
\end{equation}
where the sources incoherent with the speckles are: $\mathbf{I}^{P}$
- the intensity of the signal (planet), $\mathbf{I}^{Z}$ - the intensity
of the zodi, and $\mathbf{I}^{D}$ - the ``effective'' intensity
of the dark current (thermal electrons detected as photons). For a
space telescope pointing in a fixed direction, $\mathbf{I}^{P}$ and
$\mathbf{I}^{Z}$ are assumed to be constant and $\mathbf{I}^{D}$
is assumed to be constant, known, and uniform. In our analysis, $\mathbf{I}^{Z}$ and non-uniformities in $\mathbf{I}^{D}$ will be indistinguishable from $\mathbf{I}^{P}$ and are therefore sources of systematic errors (which could potentially be addressed given a model of the actual optical system and observation scenes).

The measured numbers of photons during frame $t$,
\begin{equation}
\mathbb{R}^{N}\ni\mathbf{y}(t)\sim\mathrm{Poisson}\left(\mathbf{I}(t)\right),\label{eq:poisson}
\end{equation}
is a vector of realizations of independent Poisson distributed variables
whose parameters are given by the elements of $\mathbf{I}(t)$. Our
goal is, given measurements $\left\{ \mathbf{y}(t)\right\} _{t=1}^{T}$ and
some model for $\mathbf{E}(t)$, to estimate $\mathbf{I}^{P}+\mathbf{I}^{Z}$
(and we will assume $\mathbf{I}^{Z}=\mathbf{0}$ for the rest of the
paper). In Eq.~(\ref{eq:poisson}), we ignored read noise since it is expected to be negligible when WFIRST is operating in ``photon-counting'' mode (\cite{harding2015technology}).

We note that errors in the Jacobian, $G^U$, would translate into errors in estimates of the field via Eq.~(\ref{eq:EOL_def}). This issue can be addressed by keeping the commands, $\mathbf{u}(t)$, approximately zero-mean and small (this requires periodically ``recalibrating'' $\mathbf{u}(0)$, see \cite{pogorelyuk2019dark}), and by obtaining more accurate estimates of the Jacobian during the dark-hole creation phase (\cite{sun2018identification}).

\subsection{\label{sub:dimensions}Reduced-Order Modeling of the Speckle Field and its Drift}

The main idea behind our method is to assume that the electric field,
$\mathbf{E}^{OL}(t)$, lies in a low dimensional subspace,
\begin{equation}
\mathbf{E}^{OL}(t)\in{\cal S}_{E}\subset\mathbb{R}^{2N}.\label{eq:E_subspace}
\end{equation}
Here, $\mathrm{dim}{\cal S}_{E}=r$ and $r-1$ is the number of Zernike polynomials (or any other basis)
required to get an accurate model of the low-order, slowly-varying WFEs in the pupil
plane. It is implied that those errors propagate linearly, hence the
increments of the hypothetical open-loop field, $\mathbf{E}^{OL}(t)-\mathbf{E}^{OL}(t-1)$,
lie in an $r-1$ dimensional subspace (the extra dimension is for
$\mathbf{E}^{OL}(0)$). Similarly, contemporary methods based on PCA
of the intensity assume that the intensity lies in a low dimensional subspace, 
\begin{equation}
\mathbf{I}^{S,OL}(t)\in{\cal S}_{I}\subset\mathbb{R}^{N},\label{eq:I_subspace}
\end{equation}
where $\mathbf{I}^{S,OL}=B\cdot \left( \mathbf{E}^{OL}\circ\mathbf{E}^{OL} \right)$. Here, ${\cal S}_{E}$ and ${\cal S}_{I}$ encompass all of the pixels, but both formulations are applicable to smaller regions of pixels in the dark hole.

The two approaches are consistent in the sense that if the field increments
satisfy Eq.~(\ref{eq:E_subspace}) then the intensity increments satisfy
Eq.~(\ref{eq:I_subspace}). However, in light of Eq.~(\ref{eq:speckles_intensity})
and the properties of the Hadamard product, 
\begin{equation}
\mathrm{dim}{\cal S}_{I}\le\binom{r+1}{2}=\frac{r(r+1)}{2}.\label{eq:dim_S_I}
\end{equation}
That is, the dimension of the intensities subspace, ${\cal S}_{I}$,
may be up to $(r+1)/2$ times larger than the dimension of the electric
fields subspace, ${\cal S}_{E}$.

We note that the dimension of ${\cal S}_{I}$ can be as small as just $4$ times larger than the dimension of ${\cal S}_{E}$ when, for instance, the WFEs happen to be spanned by the two-dimensional Fourier basis (since a product of Fourier modes is also a Fourier mode). In practice, the WFEs do not exactly lie in a low-dimensional subspace;  it therefore only makes sense to consider the ``effective'' $\mathrm{dim}{\cal S}_{I}$ which is somewhere between $4$ and $(r+1)/2$ times larger than the ``effective'' $\mathrm{dim}{\cal S}_{E}$, depending on how well the WFEs are approximated by a Fourier basis.
In any case, we claim that low-order
modelling of speckles is more naturally done in terms the electric field
rather then intensity.

To this end, we rewrite Eq.~(\ref{eq:E_subspace}) as
\begin{equation}
\mathbf{E}^{OL}(t)\in\mathrm{colsp}\left\{ G^{V}\right\} ,
\end{equation}
where ${\cal S}_{E}$ is replaced with the column space of some generally
unknown matrix $G^{V}\in\mathbb{R}^{2N\times r}$. This gives a low
order parameterization of the open loop electric field,
\begin{equation}
\mathbf{E}^{OL}(t)=G^{V}\mathbf{v}(t),\label{eq:ROM_field}
\end{equation}
with $\mathbf{v}(t)\in\mathbb{R}^{r}$. The goal of the remainder
of this section is to estimate the relatively small number of parameters
$G^{V},\left\{ \mathbf{v}(t)\right\} _{t=1}^{T}$ given measurements
and controls history. This, in turn, will allow us to estimate the planet signal, $\mathbf{I}^{P}$.

\subsection{Estimation Based on Electric Field Order Reduction}

We wish to estimate the intensity of the planet, $\mathbf{I}^{P}$,
given measurements $\left\{ \mathbf{y}(t)\right\} _{t=1}^{T}$ that
depend on the history of the unknown electric field, $\left\{ \mathbf{E}^{OL}(t)\right\} _{t=1}^{T}$,
and known DM commands, $\left\{ \mathbf{u}(t)\right\} _{t=1}^{T}$.
The proposed method has a single free parameter - the dimension of the electric
field subspace, $r$ (its implementation can be found at: \url{https://github.com/leonidprinceton/EFOR}).

First, we approximate the incoherent intensity, $\mathbf{I}^{I}=\mathbf{I}^{P}+\mathbf{I}^{D}$
(assuming $\mathbf{I}^{Z}=0$), by its conditional mean estimate,
$\tilde{\mathbf{I}}^{I}$, given by, (see Eqs.~(\ref{eq:EOL_def}),(\ref{eq:speckles_intensity})
and (\ref{eq:I_def})),
\begin{equation}
\tilde{\mathbf{I}}^{I}\left(\left\{ \mathbf{E}^{OL}(t)\right\} _{t=1}^{T}\right)=\mbox{\ensuremath{\mathrm{ramp}}}_{D}\left\{ \frac{1}{T}\underset{t=1}{\overset{T}{\sum}}\left(\mathbf{y}(t)-B\cdot\left(\left(\mathbf{E}^{OL}(t)+G^{U}\mathbf{u}(t)\right)\circ\left(\mathbf{E}^{OL}(t)+G^{U}\mathbf{u}(t)\right)\right)\right)\right\} \label{eq:I_est}
\end{equation}
where $\mbox{\ensuremath{\mathrm{ramp}}}_{D}$ is the shifted ramp
function,
\begin{equation}
\mbox{\ensuremath{\mathrm{ramp}}}_{D}\left\{ \left[I_{i}\right]_{i=1}^{N}\right\} =\left[\max\left\{ I_{i},I^{D}\right\} \right]_{i=1}^{N}.
\end{equation}
This ensures that the intensity estimate is above the dark current (which
is known).

Second, we express the probability of the measurements given the electric
fields and incoherent intensities,
\begin{equation}
p\left(\left\{ \mathbf{y}(t)\right\} _{t=1}^{T}|\left\{ \mathbf{E}^{OL}(t)\right\} _{t=1}^{T},\mathbf{I}^{I}\right)=\underset{t=1}{\overset{T}{\prod}}\underset{i=1}{\overset{N}{\prod}}\frac{\left(I_{i}^{I}+\left|E_{i}^{OL}(t)+G_{i}^{U}\mathbf{u}(t)\right|^{2}\right)^{y_{i}(t)}}{y_{i}(t)!}e^{-\left(I_{i}^{I}+\left|E_{i}^{OL}(t)+G_{i}^{U}\mathbf{u}(t)\right|^{2}\right)}.
\end{equation}
This expression stems from the fact that $y_{i}(t)$ is Poisson distributed
with parameter $I_{i}(t)=I_{i}^{I}+\left|E_{i}^{OL}(t)+G_{i}^{U}\mathbf{u}(t)\right|^{2}$.
Substituting $\tilde{\mathbf{I}}^{I}$ (the intensity estimate from
Eq.~(\ref{eq:I_est})) in place of $\mathbf{I}^{I}$, this distribution
can be approximated by a function of the field alone, 
\begin{equation}
p\left(\left\{ \mathbf{y}(t)\right\} _{t=1}^{T}|\left\{ \mathbf{E}^{OL}(t)\right\} _{t=1}^{T}\right)=\underset{t=1}{\overset{T}{\prod}}\underset{i=1}{\overset{N}{\prod}}\frac{\left(\tilde{I}_{i}^{I}+\left|E_{i}^{OL}(t)+G_{i}^{U}\mathbf{u}(t)\right|^{2}\right)^{y_{i}(t)}}{y_{i}(t)!}e^{-\left(\tilde{I}_{i}^{I}+\left|E_{i}^{OL}(t)+G_{i}^{U}\mathbf{u}(t)\right|^{2}\right)},\label{eq:p_approx}
\end{equation}
where $\tilde{I}_{i}^{I}$ is also a function of $\left\{ E_{i}^{OL}(t)\right\} _{t=1}^{T}$.

Lastly, we replace $\left\{ \mathbf{E}^{OL}(t)\right\} _{t=1}^{T}$
by its reduced-order parameterization, Eq.~(\ref{eq:ROM_field}), and
propose the maximum likelihood estimator,
\begin{alignat}{1}
\hat{G}^{V},\left\{ \hat{\mathbf{v}}(t)\right\} _{t=1}^{T}= & \underset{G^{V},\left\{ \mathbf{v}(t)\right\} _{t=1}^{T}}{\mathrm{argmax}}p\left(\left\{ \mathbf{y}(t)\right\} _{t=1}^{T}|\left\{ G^{V}\mathbf{v}(t)\right\} _{t=1}^{T}\right)\nonumber \\
\hat{\mathbf{I}}^{P}= & \tilde{\mathbf{I}}^{I}\left(\left\{ \hat{G}^{V}\hat{\mathbf{v}}(t)\right\} _{t=1}^{T}\right)-\mathbf{I}^{D}.\label{eq:estimator}
\end{alignat}
We note that the reduced rank of $\mathbf{v}(t)$ compared to $\mathbf{E}(t)$
can be seen as a regularization of the likelihood implied by Eq.~(\ref{eq:p_approx}).
In \cite{pogorelyuk2019dark}, the authors regularized the likelihood
using the prior distribution of $\mathbf{E}^{OL}(t)$ (with respect
to time), which resulted in a maximum \emph{a posteriori}
estimate.

The task of finding the estimate in Eq.~(\ref{eq:estimator}) can
be expressed as an optimization problem. We define the cost function
\begin{equation}
J\left(G^{V},\left\{ \mathbf{v}(t)\right\} _{t=1}^{T}\right)=-\log p\left(\left\{ \mathbf{y}(t)\right\} _{t=1}^{T}\left|\left\{ G^{V}\mathbf{v}(t)\right\} _{t=1}^{T}\right.\right),\label{eq:cost}
\end{equation}
where $p$ is defined via Eq.~(\ref{eq:p_approx}). The minima of
$J$ are not unique or isolated, since there are many different choices of basis vectors (columns of $G^V$) that span a given subspace. More precisely, for any invertible $W\in\mathbb{R}^{r\times r}$, we have
\begin{equation}
J\left(G^{V},\left\{ \mathbf{v}(t)\right\} _{t=1}^{T}\right)=J\left(G^{V}W,\left\{ W^{-1}\mathbf{v}(t)\right\} _{t=1}^{T}\right),
\end{equation}
so for any minimizer~$G^V$ there is a family of minimizers (parameterized by $W$) with the same cost~$J$.
In section~\ref{sec:simulation}, we find a local minimum of $J$ using a gradient descent algorithm (specifically the Adam optimizer, \cite{kingma2014adam}), in which we initialize $G^V$ as a random orthogonal matrix.  This method does not appear to have any numerical issues, even though the minima are not isolated.
Alternatively,
one could add a soft constraint term, $\left\Vert \mathrm{transpose}\left\{ G^{V}\right\} G^{V}-I\right\Vert $,
to the cost function (see for example \cite{brock2016neural}), or constrain $\left\{ \mathbf{E}^{OL}(t)\right\} _{t=1}^{T}$
to a Grassmann manifold (\cite{edelman1998geometry,townsend2016pymanopt}).

\subsection{Incorporating Reference Images}

So far we have only considered measurements, $\left\{ \mathbf{y}(t)\right\} _{t=1}^{T}$,
taken during the observation phase. However, existing reduced-order
methods incorporate prior observations by keeping a library of speckle
images which they use to compute the intensity subspace. An analogous
``library'' in our case would be a set of reference images, $\left\{ \mathbf{y}^{ref}(m)\right\} _{m=1}^{R}$,
and controls, $\left\{ \mathbf{u}^{ref}(m)\right\} _{m=1}^{R}$, applied
to the DMs when taking those images. Below, we show how to incorporate
such data to improve the incoherent intensity estimate.

To compile a library of reference PSFs to be used with KLIP, WFIRST is proposing to periodically chop from the target star to a brighter reference star to obtain new reference PSFs and to occasionaly reset the DMs.
Although the reference star is
usually significantly brighter than the target star, there is no conceptual
difference between the measurements. Therefore, we treat the two types
of data identically and compute an estimate of the speckles drift subspace,
$\mathrm{\mathrm{colsp\left\{ G^{V}\right\} }}$, based on both. This is contrary to PCA approaches which do not employ DMs to allow ``detecting'' incoherent sources in the library of reference images (and therefore target images cannot be used as part of that library).

For simplicity, we assume that in addition to the measurements of the
target star, $\left\{ \mathbf{y}(t)\right\} _{t=1}^{T}$, we have a
set of images corresponding to a single reference star, $\left\{ \mathbf{y}^{ref}(m)\right\} _{m=1}^{R}$.
The controls Jacobian, $G^{U,ref}$, depends on the brightness and
spectrum of the reference star and therefore differs from the Jacobian
of the target star, $G^{U}$. The parameterizations of the electric
field in Eq.~(\ref{eq:ROM_field}) may be shifted by some known $\mathbf{E}_{0}^{OL,ref}$,
if the nominal controls differ between the two observations. The speckles
subspace, on the other hand, is a property of the optical system;
 it is thus shared between the observations ($G^{V,ref}=G^{V}$).

A reduced-order parameterization of the open-loop field of the reference
star is (see Eq.~(\ref{eq:ROM_field})), 
\begin{equation}
\mathbf{E}^{OL,ref}(m)=\mathbf{E}_{0}^{OL,ref}+G^{V}\mathbf{v}^{ref}(m).
\end{equation}
Likewise, the conditional incoherent intensity estimate in Eq.~(\ref{eq:I_est})
becomes 
\begin{equation}
\tilde{\mathbf{I}}^{I,ref}\left(\left\{ \mathbf{E}^{OL,ref}(m)\right\} _{m=1}^{R}\right)=\mbox{\ensuremath{\mathrm{ramp}}}_{D}\left\{ \frac{1}{R}\underset{m=1}{\overset{R}{\sum}}\left(\mathbf{y}^{ref}(m)-B\cdot\left(\mathbf{E}^{OL,ref}(m)+G^{U,ref}\mathbf{u}^{ref}(m)\right)^{\circ2}\right)\right\} ,
\end{equation}
and the cost function for the reference star (defined via Eq.~(\ref{eq:p_approx})
and $\tilde{\mathbf{I}}^{I,ref}$) is
\begin{equation}
J^{ref}\left(G^{V},\left\{ \mathbf{v}^{ref}(m)\right\} _{m=1}^{R}\right)=-\log p\left(\left\{ \mathbf{y}^{ref}(m)\right\} _{m=1}^{R}\left|\left\{ \mathbf{E}_{0}^{OL,ref}+G^{V}\mathbf{v}^{ref}(m)\right\} _{m=1}^{R}\right.\right).
\end{equation}

We may now find the incoherent intensity estimate of the target star,
$\hat{\mathbf{I}}^{P}$, by optimizing
\begin{equation}
J^{tot}\left(G^{V},\left\{ \mathbf{v}^{ref}(m)\right\} _{m=1}^{R},\left\{ \mathbf{v}(t)\right\} _{t=1}^{T}\right)=J+J^{ref},\label{eq:J_tot}
\end{equation}
and then using Eq.~(\ref{eq:estimator}). Note that measurements corresponding
to higher intensity (higher $y_{i}$ or $y_{i}^{ref}$), have proportionally
larger ``weights'' in $J^{tot}$ due to the power term in Eq.~(\ref{eq:p_approx}).
This implies that the data gathered from the brighter reference star
plays a larger role in determining the speckle drift modes, $G^{V}$,
because they carry more information (per image).

Since Eq.~(\ref{eq:J_tot}) treats the measurements from various sources
identically, it can be extended to incorporate data from multiple
targets (and possibly reference) stars,
\begin{equation}
J^{tot}=J^{tar,1}+J^{tar,2}+\cdots+\left(J^{ref,1}+J^{ref,1}+\cdots\right).\label{eq:J_multiple_targets}
\end{equation}
This will arguably give better characterization of
the drift, $G^{V}$, and consequently result in better estimates of
the incoherent intensities for the target stars, $\left\{ \hat{\mathbf{I}}^{P,1},\hat{\mathbf{I}}^{P,2},\cdots \right\}$. Given enough measurements,
one might even be able to improve accuracy of the Jacobian estimate by adding
appropriate terms to Eq.~(\ref{eq:J_tot}) (similarly to the real-time
approach for combined estimation of the electric field and the Jacobian
proposed in \cite{sun2018identification}).

\section{\label{sec:simulation}Numerical Simulation}

This section illustrates the Electric Field Order Reduction (EFOR)
method for estimating the incoherent intensity given a history of
photon counts (images), control inputs, and, possibly, a set of reference
images (the data and code for generating the results below are given
at: \url{https://github.com/leonidprinceton/EFOR}). The data was simulated with the Fast Linearized Coronagraph
Optimizer (FALCO) (\cite{riggs2018fast}) model of the WFIRST hybrid
Lyot coronagraph.

\subsection{Simulation Details}

\begin{figure}
\plotone{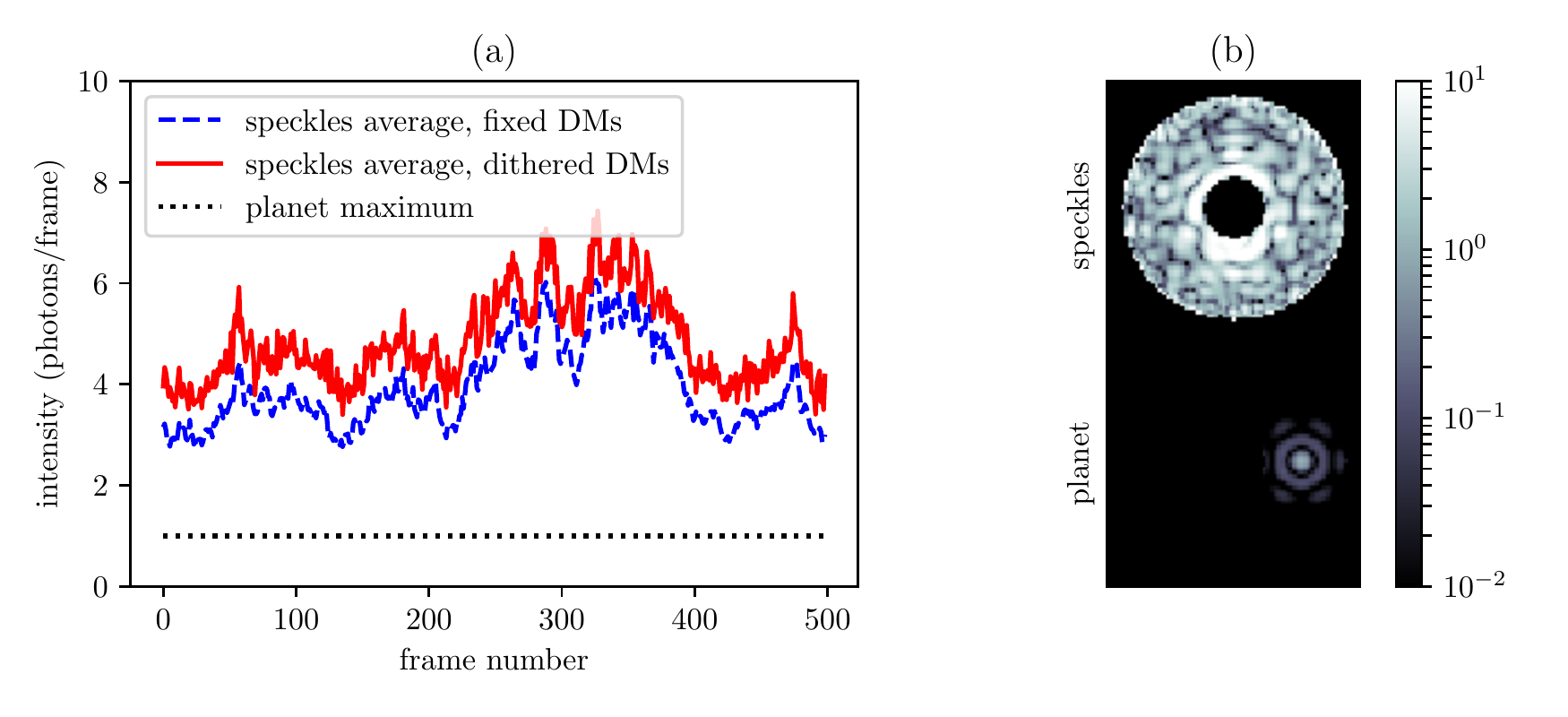}
\caption{\label{fig:speckles}(a) Average intensity of the speckles in the
dark hole (with and without dithering of the deformable mirrors) as
it varies in time due to WFE drift. The maximum intensity of the
planet is an order of magnitude below the speckle floor. (b) Side by
side comparison of the speckles and the planet on a logarithmic scale. The data was obtained
using a model of the WFIRST hybrid Lyot coronagraph implemented in
FALCO (\cite{riggs2018fast}).}
\end{figure}

The hybrid Lyot coronagraph starts at a relatively low contrast due to model/system imperfections. Therefore, as a precursor to the observation stage, the EFC (\cite{give2007electric})
algorithm was used to create a dark hole in a ring between $3$ and
$9$ $\lambda/D$ with a nominal contrast of $4.2\cdot10^{-10}$
at $\lambda=546\:\mathrm{nm}$ (to simplify the analysis, we discarded all electric fields of longer wavelengths that were simulated by FALCO). At this point, we introduced wavefront
drift through independent random increments of the first $21$ Zernike polynomials,
\begin{equation}
\Delta z_{p}^{p-2j}(t)\sim{\cal N}\left(0,\frac{0.1\:\mathrm{nm}}{p^{2}\cdot\lambda}\right),\;0\le j\le p,
\end{equation}
where $p$ is the order of the polynomial, $j$ is its azimuthal degree
and $\Delta z_{p}^{p-2j}$ is its increment over one $100\:\mathrm{sec}$
frame. To make the comparison with PCA based methods (\cite{soummer2012detection,amara2012pynpoint}) easier, we chose a Markov drift model in which the Zernike coefficients $z_{p}^{p-2i}(t)$
fluctuate but remain bounded at all times $t$,
\begin{equation}
z_{p}^{p-2j}(t)=0.99\cdot z_{p}^{p-2j}(t-1)+\Delta z_{p}^{p-2j}(t).\label{eq:zernike_drift}
\end{equation}
After creating the dark hole, but before ``making'' the observations,
the Zenike coefficients where advanced $1000$ time steps via Eq.~(\ref{eq:zernike_drift}). Another potential source of high-order wavefront error is the time-dependent drift of the DM actuators themselves (J. T. Trauger, personal communication, 2019). Although not addressed here, its influence on the electric field would be linear and manifest itself through a low-rank approximation of the control Jacobian ($G^U$ instead of $G^V$ in Eq.~(\ref{eq:ROM_field})). Contrary to what one might expect, the ``effective'' rank of the Jacobian is far lower than the number of actuators, the latter being usually large to prevent actuator over-saturation.

The simulated data consisted of $500$ frames of the target star (with
an exoplanet) and $1500$ frames of the reference star, which is more than sufficient for PCA purposes. 
The measurements were ``taken'' once with fixed DMs and once with DM
dithering, while keeping the drift history, $z_{p}^{p-2i}(t)$, the
same. The average intensity of the residual light was $1.2\:\mathrm{\frac{photons}{frame}}$
in a perfect dark hole, $4.8\:\mathrm{\frac{photons}{frame}}$ with
simulated WFEs (time average), and $5.7\:\mathrm{\frac{photons}{frame}}$
with both WFEs and DM dithering. When pointing at the reference start,
all speckle intensities where $16$ times higher. The maximum intensity
of the planet was $1.0\:\mathrm{\frac{photons}{frame}}$ (see comparison
in Fig.~\ref{fig:speckles}(b)) and the effective intensity of the
dark current was $0.25\:\mathrm{\frac{photons}{frame}}$.

The total number of pixels in the dark hole (for the single wavelength
considered) was $N=2608$. The history of the average dark hole intensity
when observing the target star is shown in Fig.~\ref{fig:speckles}(a),
with and without DM dithering (the dithering for each actuator was sampled from a normal distribution with $0.01$V standard deviation around their nominal values which had an average magnitude of $10$V). At each time frame, the number of photons
detected at each pixel was sampled from a Poisson distribution based
on the intensity of that pixel (Eq.~(\ref{eq:poisson})). We note
 that instead of just dithering the DMs, one could also apply an EFC control law to mitigate the effects of time-varying speckles (\cite{pogorelyuk2019dark}). This closed-loop approach would have resulted in higher contrast,
lower shot noise, and a better post-processing factor; we plan to analyze the combined performance of EFOR and dark-hole maintenance in a future paper.

\subsection{Results and Discussion}

\begin{figure}
\plotone{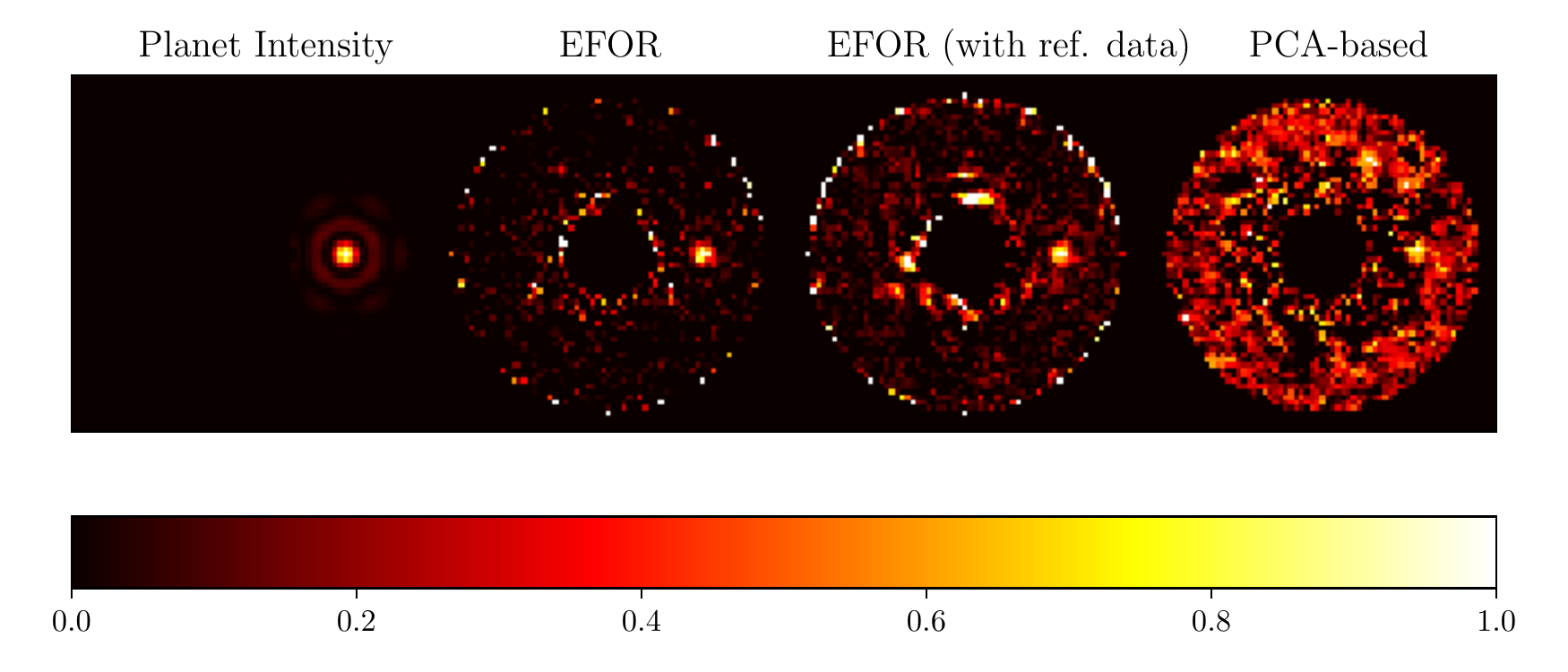}
\caption{\label{fig:I_estimates}Planet intensity and its estimates by newly
proposed Electric Field Order Reduction (EFOR) method (with and without
a library of reference images) and by Principal Component Analysis
(PCA; see \cite{amara2012pynpoint}). The images used by EFOR were
simulated with an additional dithering of the DMs, but the underlying
wavefront errors were the same.}
\end{figure}

Fig.~\ref{fig:I_estimates} shows the best estimates of the incoherent
intensity obtained by PCA and the EFOR method introduced in sec.~\ref{sec:EFOR}.
The PCA estimates were computed based on measurements taken with fixed
DMs. The reference images were used to construct a basis for
speckle intensities (via Singular Value Decomposition) across \emph{all}
pixels (although splitting the domain into smaller regions could reduce
the optimal number of modes, see \cite{soummer2012detection,fergus2014s4}).
The two EFOR estimates were obtained from measurements taken while
dithering the DMs: one corresponds to fitting just the observations
of the target star, and the other to fitting the observations to both
the target and the reference stars (Eq.~(\ref{eq:cost}) vs.~(\ref{eq:J_tot})).
The latter estimate is slightly more accurate, illustrating the advantage of incorporating multiple observations. However, once one has a good estimate of the speckle basis, $\mathrm{\mathrm{colsp\left\{ G^{V}\right\} }}$, additional reference images provide diminishing improvements in accuracy (similarly, PCA becomes insensitive to the number of reference images beyond a certain threshold).

\begin{figure}
\plotone{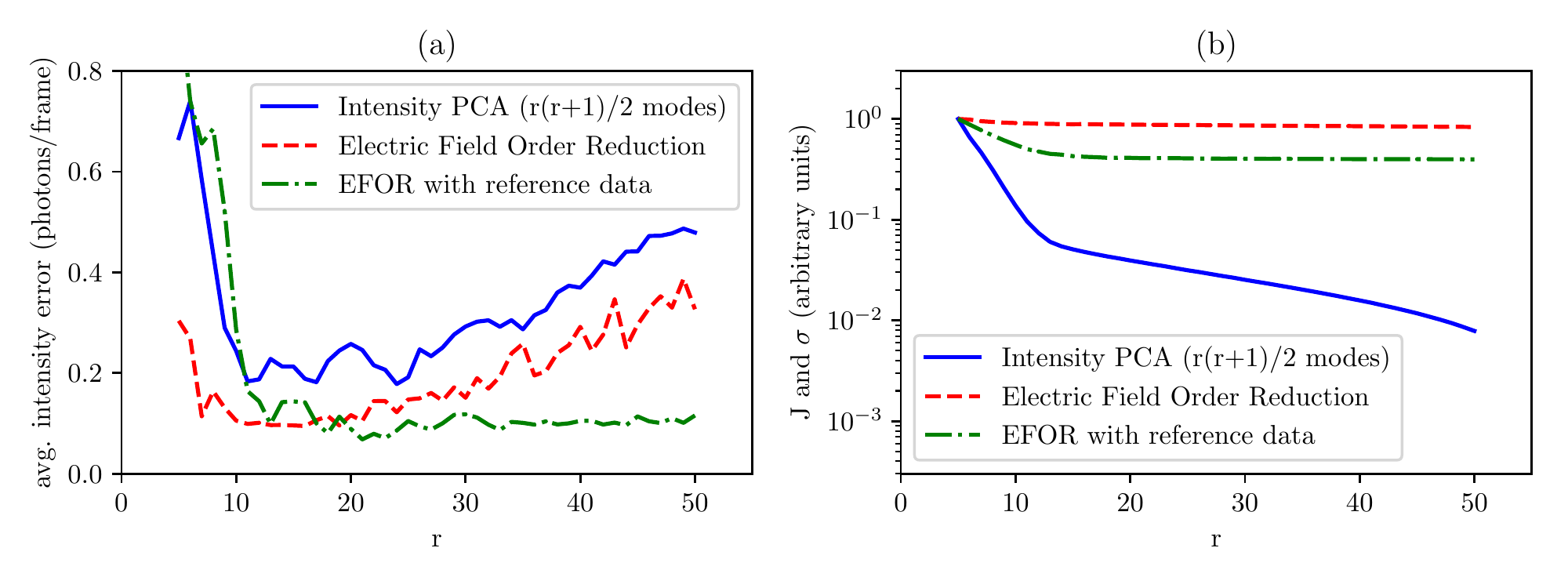}
\caption{\label{fig:I_errors}(a) Average errors in intensity estimates in
planet's half-max region ($I_{i}^{P}>0.5\left\Vert \mathbf{I}^{P}\right\Vert _{\infty}$),
as a function of the number of electric field modes $r$ and the number
of intensity modes $\frac{r(r+1)}{2}$. EFOR outperforms
PCA by a factor of about $2.5$ when reference data is present. In the
absence of reference data, EFOR tends to overfit the data 
when $r>r_{true}=22$. (b) Arbitrarily scaled residual errors of EFOR
(the cost function in Eqs.~(\ref{eq:cost}) and (\ref{eq:J_tot}))
and PCA (truncated singular value $\sigma$). The PCA and EFOR curves
that are based on reference images suggest that the number of meaningful
electric field modes is around $r=15$.}
\end{figure}

To make the comparison between the methods quantitative, consider the
average error in the incoherent intensity estimate, $\mathbf{\hat{I}}^{P}$,
in a region where the planet's intensity is above its half-maximum,
\begin{equation}
\mathrm{err}\left(\mathbf{\hat{I}}^{P}\right)\equiv\mathrm{avg}\left\{ \left.\left|\hat{I}_{i}^{P}-I_{i}^{P}\right|\right|1\le i\le N,\:I_{i}^{P}>0.5\left\Vert \mathbf{I}^{P}\right\Vert _{\infty}\right\} .\label{eq:err_def}
\end{equation}
In Fig.~\ref{fig:I_errors}(a), the errors in EFOR estimates are plotted
as a function of the number of modes, $r$, while the errors in the
PCA estimates are plotted as a function of $\frac{r(r+1)}{2}$ (in
light of Eq.~(\ref{eq:dim_S_I})). We indeed see that the errors of
EFOR with reference data flatten around $r_{true}=22$ modes, while
PCA reaches its peak accuracy in the region between $66$ and $300$ modes (corresponding to roughly $4r_{true}$ and $\frac{r_{true}(r_{true}+1)}{2}$, see discussion after Eq.~(\ref{eq:dim_S_I})). EFOR without
reference data, on the other hand, reaches its peak accuracy at around
$r=15$ modes after which it overfits the data.

The lowest error, as defined by Eq.~(\ref{eq:err_def}), is achieved
by EFOR with reference data. The minimum error of EFOR with data
from just the target start is $1.4$ times larger, while the lowest
PCA error is $2.5$ times larger. These errors can be further reduced
by incorporating more observations or employing a closed loop dark
hole maintenance scheme (\cite{pogorelyuk2019dark}) to reduce the shot
noise.

Finally, Fig.~\ref{fig:I_errors}(b) shows the cost function $J$ of
the EFOR methods and the truncated singular value of the PCA method
as a function or $r$ and $\frac{r(r+1)}{2}$ respectively. The steep
decrease in the EFOR's cost function (with reference
data) stops at $r=15$, suggesting that this
is the number of ``important'' WFE modes. One would therefore expect the steep decrease in the singular values to stop somewhere between $4r=60$ and $\frac{r(r+1)}{2}=120$ PCA modes, and it indeed stops at around $\frac{12\cdot13}{2}=78$ modes. In the case of a single observation sequence, the cost function of EFOR decreases very slowly (red dashed line on Fig.~\ref{fig:I_errors}(b)) and doesn't provide an indication of the ``correct'' number of modes, $r$. However, instead of using reference images, one could incorporate data from multiple targets to get a better estimate of the speckle field basis, $G^{V}$ (see discussion around Eq.~(\ref{eq:J_multiple_targets})). Alternatively, one could choose the $r$ that maximizes some detection criterion based on the PSF of the planet/halo (e.g. a matched filter, \cite{kasdin2006linear}, or a Hotelling observer, \cite{caucci2007application}).

\section{Conclusions}

In this work we assumed that the slow evolution of wavefront errors in
space based coronagraphs can be well approximated by a small number
of modes of the electric field in the pupil plane and that these errors are
linearly propagated to the image plane. We  then showed that a
low-dimensional approximation of the speckles requires an order of magnitude
fewer electric field modes than intensity modes. Consequently, we introduced
a reduced-order algorithm for estimating the history of the electric
field and the incoherent intensity (which includes the planet signal).
Unlike existing methods (e.g. KLIP),
our algorithm relies on small actuations (dithering) of the deformable
mirrors during the observation phase as well as an accurate estimate
of the controls influence matrix (the Jacobian). This introduces the
phase diversity required for estimating the electric field, rather
than the intensities of the speckles.

We illustrated that since the newly proposed method employs significantly
fewer parameters, it has the potential to make substantially more accurate
estimates than intensity-based PCA, even without a library of reference
images. However, our method does allow for seamlessly incorporating
reference data as well as other observations. The resulting scheme
searches for an ``optimal'' basis for speckle field modes across
all measurements taken by a given instrument.

More importantly, the proposed algorithm uniquely incorporates the
history of control inputs employed during the observation phase. We
speculate that combining it with a closed-loop dark hole maintenance
scheme will decrease the shot noise and increase the post-processing
factor even further. This direction, together with an analysis of the
dithering strategy, are subjects of future research.

\section*{Acknowledgements}
This material is based upon work supported by the Army Research Office, award number W911NF-17-1-0512. We would like to thank Neil T. Zimmerman for providing technical details for the proposed observation sequence of the WFIRST telescope. We would also like to thank the reviewer of the manuscript for pointing out the lower bound on the number of modes necessary to describe a set of intensity fields in section~\ref{sub:dimensions}.

\bibliography{EFOR}

\end{document}